\documentclass[runningheads]{llncs}

\usepackage{tabularray}
\usepackage{color}
\usepackage{bbding}
\usepackage{subcaption,graphicx}
\usepackage{subcaption}
\usepackage{float}
\usepackage{eccv}

\usepackage{eccvabbrv}
\usepackage{wrapfig}
\usepackage{graphicx}
\usepackage{booktabs}

\usepackage[accsupp]{axessibility}  % Improves PDF readability for those with disabilities.

\usepackage[pagebackref,breaklinks,colorlinks,citecolor=eccvblue]{hyperref}
\usepackage{orcidlink}
\usepackage{multirow}
\usepackage{booktabs}
\usepackage{colortbl}
\usepackage{makecell}                 % 三线表-竖线

\begin{document}

\title{Learning Exhaustive Correlation for Spectral Super-Resolution: Where Spatial-Spectral Attention Meets Linear Dependence}
\titlerunning{Learning Exhaustive Correlation for Spectral Super-Resolution}

% Second Author\inst{2,3}\orcidlink{1111-2222-3333-4444} \and
% Third Author\inst{3}\orcidlink{2222--3333-4444-5555}}
\author{Hongyuan Wang\inst{1} \and
Lizhi Wang\inst{1}\and
Jiang Xu\inst{1}\\Chang Chen\inst{2} \and Xue Hu\inst{2} \and Fenglong Song\inst{2} \and Youliang Yan\inst{2}}
\authorrunning{Hongyuan Wang et al.}

% Springer Heidelberg, Tiergartenstr.~17, 69121 Heidelberg, Germany
\institute{Beijing Institute of Technology \and
Huawei Noah's Ark Lab
}
\maketitle

\begin{abstract}
  Spectral super-resolution that aims to recover hyperspectral image (HSI) from easily obtainable RGB image has drawn increasing interest in the field of computational photography. The crucial aspect of spectral super-resolution lies in exploiting the correlation within HSIs. However, two types of bottlenecks in existing Transformers limit performance improvement and practical applications. First, existing Transformers often separately emphasize either spatial-wise or spectral-wise correlation, disrupting the 3D features of HSI and hindering the exploitation of unified spatial-spectral correlation. Second, existing self-attention mechanism always establishes full-rank correlation matrix by learning the correlation between pairs of tokens, leading to its inability to describe linear dependence widely existing in HSI among multiple tokens. To address these issues, we propose a novel Exhaustive Correlation Transformer (ECT) for spectral super-resolution. First, we propose a Spectral-wise Discontinuous 3D (SD3D) splitting strategy, which models unified spatial-spectral correlation by integrating spatial-wise continuous splitting strategy and spectral-wise discontinuous splitting strategy. Second, we propose a Dynamic Low-Rank Mapping (DLRM) model, which captures linear dependence among multiple tokens through a dynamically calculated low-rank dependence map. By integrating unified spatial-spectral attention and linear dependence, our ECT can model exhaustive correlation within HSI. The experimental results on both simulated and real data indicate that our method achieves state-of-the-art performance. Codes and pretrained models will be available later.
  \end{abstract}

\section{Introduction}
\label{sec:intro}

Hyperspectral image (HSI) consists of multiple channels, with each channel representing the response in a specific spectral band. In comparison to the 3-channel RGB image, HSI excels in capturing detailed spectral information from a scene. Owing to this advantage, HSI finds extensive applications in image classification~\cite{camps2005kernel,hu2015deep,melgani2004classification}, object detection~\cite{liang2013salient}, face recognition~\cite{zhang2021hyperspectral}, and more. \begin{wrapfigure}{t}{0.51\textwidth}
	\vspace{-4.3mm}
	\begin{center}
		\includegraphics[width=0.51\textwidth]{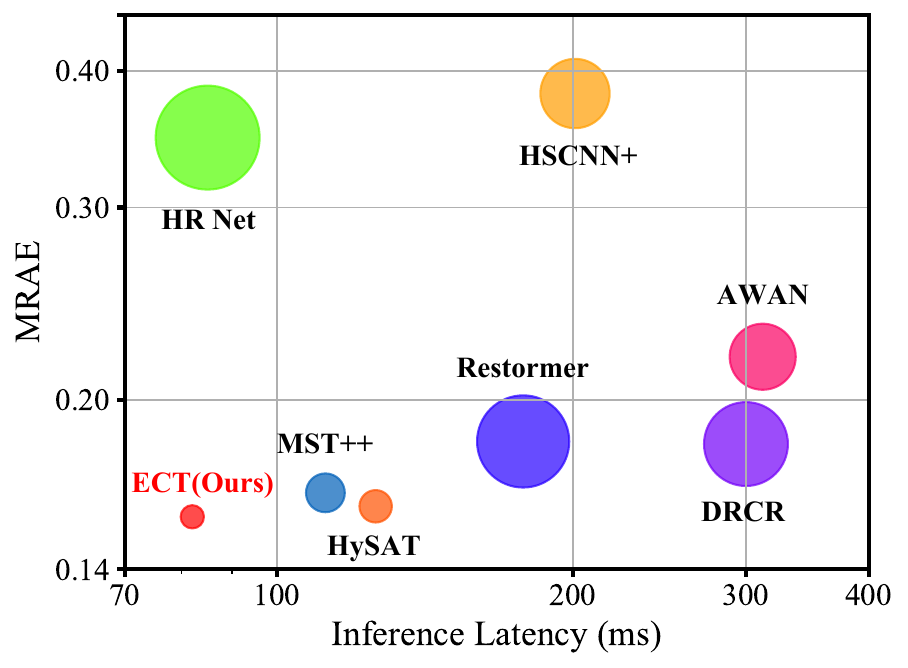}
	\end{center}
	\vspace{-6mm}
   \caption{Comparisons of MRAE, inference latency, and parameters on the NTIRE 2022 dataset are presented. The circle radius represents 
the number of parameters. }
	\vspace{-8mm}
	\label{fig:bubble}
\end{wrapfigure} However, acquiring 3D HSI with 2D sensors is challenging due to the mismatch of dimensions. Traditional scanning-based methods typically require 
multiple exposures to capture a full HSI, which is disadvantageous for dynamic and rapidly changing scenes.

To address this issue, researchers have designed snapshot compressive imaging (SCI) systems with customized optical modulation and reconstruction algorithms, enabling snapshot acquisition of HSI~\cite{yuan2021snapshot, xiong2017hscnn, wang2020dnu, cai2022mask, li2023pixel}. However, these methods are often expensive and bulky in system implementation. Consequently, the task of HSI reconstruction from the easily obtainable RGB image~\cite{zhang2022survey,he2023spectral}, known as spectral super-resolution, has emerged as a popular solution with the advantages of being cheap and lightweight.

The crucial aspect of spectral super-resolution lies in exploiting correlations within HSI. Early research utilizes sparse coding~\cite{arad2016sparse} or low-rank representation~\cite{xue2021spatial} for spectral super-resolution. However, these methods often suffer from limited expressive power and generalization ability, thus failing to achieve satisfactory results.

With the increasing computing power, learning-based methods have made significant progress in recent years and have become the mainstream solution for spectral super-resolution. Currently, Transformers~\cite{cai2022mst++,wang2023learning_lite} have attained the state-of-the-art performance for spectral super-resolution by leveraging spectral-wise correlation through a spectral-wise self-attention mechanism. However, two types of bottlenecks exist that limit performance improvement and practical applications. First, existing Transformers predominantly focus on spectral-wise correlation while overlooking spatial-wise correlation in spectral super-resolution. Some works in other tasks~\cite{li2023spatial,chen2023activating,wang2022s,liu2023unified} attempt to model both spectral-wise and spatial-wise correlations together but often utilize separate network modules. The neglect and separation undermine the 3D nature of HSI and hinder the exploitation of unified spectral-spatial correlation. 
Second, existing spectral-wise self-attention mechanism always captures the full-rank correlation matrix by learning the correlation between pairs of spectral bands, \ie tokens, in the Transformer. These characteristics result in the inability to establish linear dependence widely existing in HSI among multiple tokens.

In this paper, we propose a novel Exhaustive Correlation Transformer (ECT) to model the unified spatial-spectral correlation and linear dependence, both of which we believe are crucial for spectral super-resolution. The first motivation behind our method stems from the spatial-spectral similarity in HSI. Thus, we propose a Spectral-wise Discontinuous 3D (SD3D) splitting strategy to simultaneously model unified attention along the spectral and spatial dimensions. The SD3D splitting strategy contains continuous splitting in the spatial dimension and discontinuous splitting in the spectral dimension, allowing for an effective focus on spectral-wise non-local features without disrupting the continuous structure in the spatial dimension. 

The second motivation behind our methods arises from the information redundancy in HSI and its low-rank characteristic~\cite{fu2016exploiting,dian2017hyperspectral,zhang2019computational}. Thus, we propose a Dynamic Low-Rank Mapping (DLRM) module to capture the linear dependence among multiple tokens.  The DLRM module simultaneously interacts among multiple tokens and maps them into a low-rank space, thereby learning a low-rank dependence map.
By integrating unified spatial-spectral attention and linear dependence, our ECT can model the exhaustive correlation within HSI and achieves state-of-the-art in extensive experiments on simulated and real data.

Our contributions are summarized as follows:
\begin{itemize}

\item We propose a novel Exhaustive Correlation Transformer (ECT) to model unified spatial-spectral correlation and linear dependence for spectral super-resolution.

\item We propose a Spectral-wise Discontinuous 3D (SD3D) splitting strategy to exploit the unified spatial-spectral correlation within HSI by concurrently adopting spatial-wise continuous and spectral-wise discontinuous splitting.

\item We propose a Dynamic Low-Rank Mapping (DLRM) module to model the linear dependence within HSI by dynamically calculating a low-rank dependence map among multiple tokens.

\item The experimental results indicate that our method achieves state-of-the-art performance on both simulated and real data, with the lowest error achieved under the smallest number of parameters and the lowest inference latency.

\end{itemize}

\section{Related Work}
\label{sec:related_work}

\subsection{Spectral Reconstruction}

HSI acquisition is typically carried out using push-broom cameras, which is time-consuming and challenging to capture dynamic or rapidly changing scenes. To address this issue, coded aperture snapshot spectral imaging (CASSI) systems have been widely used, generating 2D measurements~\cite{yuan2021snapshot,llull2013coded}, which are then processed through a series of reconstruction algorithms~\cite{xiong2017hscnn,wang2019hyperspectral,wang2020dnu,cai2022coarse,cai2022degradation,cai2023binarized,yuan2020plug} to obtain HSI. 

However, CASSI systems are often expensive. Reconstructing HSIs from RGB images is a cost-effective alternative. Arad et al.~\cite{arad2016sparse} employ sparse coding for spectral super-resolution, while Aeschbacher et al.~\cite{aeschbacher2017defense} use shallow learning models and achieve improved results. Due to the presence of substantial redundant information in HSI, a low-rank prior is critical for spectral reconstruction. There are several spectral reconstruction works~\cite{zhang2021learning,zhang2019computational,fu2016exploiting,xue2021spatial,dian2017hyperspectral,dian2019hyperspectral,dian2019learning} inspired by the low-rank prior. Recently, Three spectral super-resolution challenges~\cite{arad2016sparse,arad2020ntire,arad2022ntire} are held and significantly inspire the research. With the development of deep learning, convolutional neural networks are widely used in the spectral super-resolution task. Shi et al.~\cite{shi2018hscnn+} propose a convolutional neural network for spectral super-resolution, which win the NTIRE 2018 Challenge on Spectral Reconstruction from RGB Images. Li et al.~\cite{li2020adaptive,li2022drcr} introduce the channel attention mechanism into the convolutional neural network to improve the performance. Thanks to dynamic weights and long-range correlation modeling, Cai et al.~\cite{cai2022mst++} are the first to introduce Transformers into the field of spectral super-resolution for modeling spectral-wise correlation and win first place in the NTIRE 2022 challenge~\cite{arad2022ntire}. Wang et al.~\cite{wang2023learning_lite} further improve the modeling ability of spectral-wise correlation and achieve SOTA performance recently. Though recent methods based on Transformer achieve remarkable performance improvements, they only focus on modeling spectral-wise correlation while ignoring spatial-wise correlation. Moreover, both of them neglect the critical low-rank characteristic of HSI.

% With the popularity of deep learning methods, convolutional neural networks have been widely applied in spectral super-resolution~\cite{shi2018hscnn+,li2020adaptive,stiebel2018reconstructing}. Thanks to dynamic weights and long-range correlation modeling, Cai et al.~\cite{cai2022mst++} were the first to introduce Transformers into the field of spectral super-resolution and won the first place in the NTIRE 2022 challenge~\cite{arad2022ntire}.

\subsection{Transformer Model}

In the field of NLP, to capture long-range dependencies and enable parallel processing, Vaswani et al.~\cite{vaswani2017attention} introduced the Transformer model based on the self-attention mechanism. Thanks to its capability to capture long-range dependencies, global receptive fields, and dynamic weight computation, Dosovitskiy et al.~\cite{dosovitskiy2020image} applied the Transformer to image classification, achieving outstanding results. The Transformer architecture has found widespread use in high-level computer vision tasks such as image classification~\cite{ali2021xcit,yu2022metaformer,chen2021crossvit,li2022efficientformer,liu2021swin}, semantic segmentation~\cite{strudel2021segmenter,xie2021segformer,lu2021simpler}, and object detection~\cite{carion2020end,ding2019learning}. Furthermore, in low-level computer vision tasks, Transformer-based models have also demonstrated remarkable performance in tasks like image super-resolution~\cite{chen2023activating,liang2021swinir,zou2022self,li2023efficient}, deraining~\cite{zamir2022restormer,xiao2022image,wang2022uformer}, and denoising~\cite{zamir2022restormer,liang2021swinir,li2023efficient,zhao2023comprehensive,wang2022uformer}. Transformers that leverage the self-attention mechanism can capture long-range correlations between Transformer tokens through dot-product similarity calculations and adaptively fuse tokens based on these correlations, offering strong expressive power.

From the perspective of feature maps, token splitting occurs in the spatial~\cite{dosovitskiy2020image,liu2021swin} or spectral dimensions\cite{ali2021xcit,zamir2022restormer,cai2022mst++}, allowing for modeling the relationships between pixels or patches or between channels. While there are some efforts to combine these two types of Transformers to model spatial and spectral correlations~\cite{li2023spatial,chen2023activating,wang2022s,liu2023unified}, most of these works directly treat spatial and spectral Transformers as separate modules, which destroys the 3D nature and can not fully exploit the unified correlations.
% between spatial and spectral dimensions. However, capturing this kind of correlation is crucial for HSI reconstruction.

\begin{figure}[t]
   \centering
    % \includegraphics{sec/figs/fig_bubble.pdf}
 \includegraphics[width=1.0\linewidth]{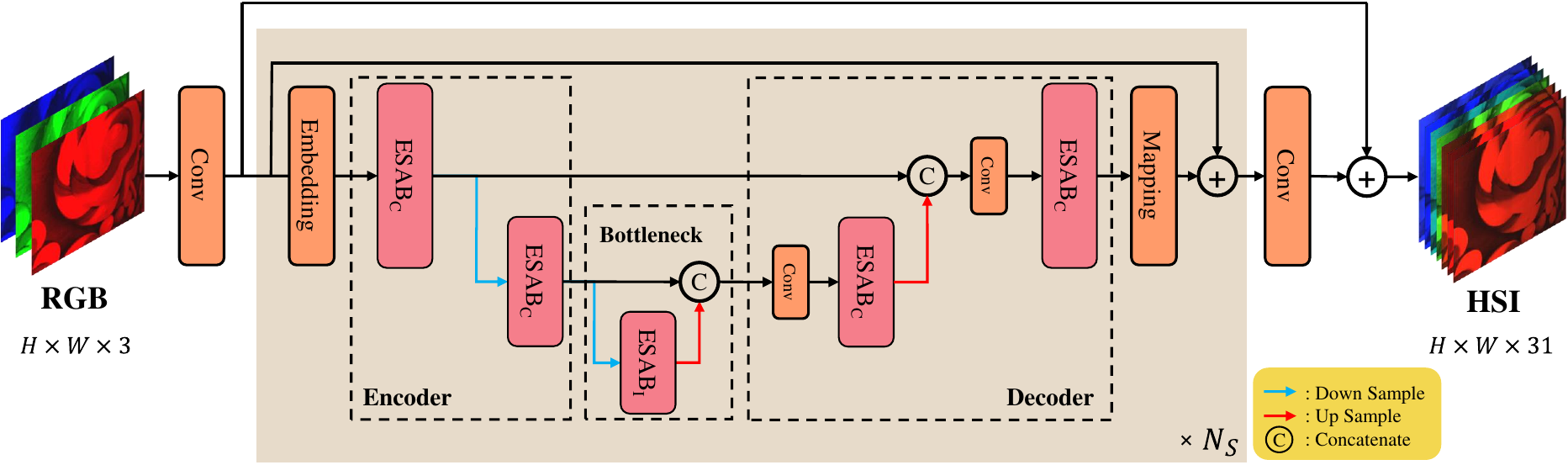}
    \caption{The macro design of Exhaustive Correlation Transformer (ECT).} %\TODO{Delete 'ECT' in the figure}
    \vspace{-1em}
    \label{fig:ect}
 \end{figure}

\section{Method}

In this paper, we propose an Exhaustive Correlation Transformer (ECT), which can model unified spatial-spectral attention and linear dependence simultaneously. In this section, we first introduce the macro design of ECT. Then, we delve into the micro design within ECT.
%In this subsection, we detail the crucial model in ECT, the Exhaustive Self-Attention (ESA), and two important models in ESA, the Unified Spatial-Spectral self-Attention (USSA) and the Dynamic Low-Rank Mapping (DLRM).

\subsection{Macro Design}
% 如图1所示，我们提出Exhaustive Correlation Transformer (ECT)。整体网络结构是一个多阶段U形结构，由于上采样和下采样的存在，这种结构可以在减少计算量的同时捕获多尺度空间特征。对于一个3通道的RGB输入，通过一个3×3的卷积将其扩展到31通道，随后送入Ns个U形模块中。每个U形模块包括Embedding、Encoder、bottleneck、decoder和Mapping。embedding和Mapping用3×3卷积实现，分别在输入侧将特征图的通道扩展到32，在输出侧将通道缩减回31.Encoder、Decoder的主体为Cross Exhaustive Self-Attention Block (ESAB)，分别用4×4，stride=2的卷积进行下采样、2×2，stride=2的转置卷积进行上采样。Bottleneck包括一层Inter Exhaustive Self-Attention Block (ESAB)。 特征图的空间分辨率在下采样后变为1/4，通道变为2倍。多头注意力的头数随着通道数的增减成倍增减，初始值为1.Encoder和Decoder之间有残差连接，保留更多的输入信息用于重建。此外添加了一个长距离残差用于稳定训练。ESAB用于建模Token之间的Correlation，ESAB用于建模Token内部的Correlation，分别建模non Local与Local Unified Spatial-spectral correlation。
We propose an Exhaustive Correlation Transformer (ECT) for spectral super-resolution. The overall network employs a multi-stage U-shaped architecture, as shown in Figure~\ref{fig:ect}. For a 3-channel RGB input, it is expanded to 31 channels using a 3$\times$3 convolution and then processed through $N_s$ U-shaped modules. Each U-shaped module consists of Embedding, Encoder, Bottleneck, Decoder, and Mapping components. Embedding and Mapping are implemented with 3$\times$3 convolutions, expanding the channel dimensions to 32 on the input side and reducing them back to 31 on the output side.
The main components of the Encoder and Decoder are the Cross Exhaustive Self-Attention Blocks (ESAB$\rm {_C}$), utilizing 4$\times$4 convolutions with a stride of 2 for downsampling and 2$\times$2 transpose convolutions with a stride of 2 for upsampling. The Bottleneck includes a layer of Inter Exhaustive Self-Attention Block (ESAB$\rm {_I}$). ESAB$\rm {_C}$ is employed to model the correlations between tokens, while ESAB$\rm {_I}$ is used to model the correlations within tokens. ESAB$\rm {_C}$ can model spatial-wise non-local and global-aware spectral-wise local correlation, while ESAB$\rm {_I}$ can model spatial-wise local and spectral-wise non-local correlation. The spatial resolution of the feature map becomes 1/4 after downsampling, while the channel doubles. The number of attention heads scales with the channel changes.
Residual connections exist between the Encoder and Decoder, retaining more input information for reconstruction. Furthermore, a long-range residual connection is added to stabilize the training. 

\begin{figure}[t]
   \centering
    % \includegraphics{sec/figs/fig_bubble.pdf}
 \includegraphics[width=1.0\linewidth]{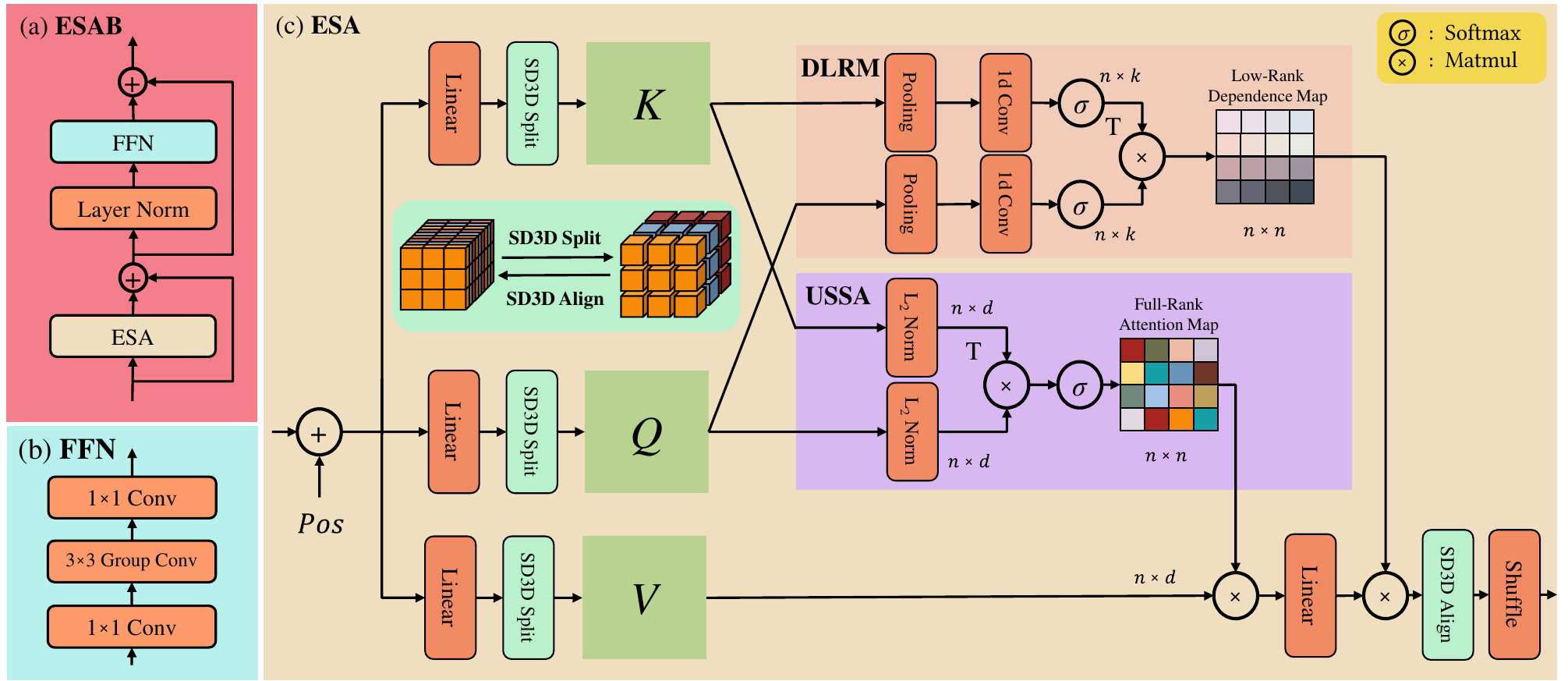}
    \caption{Micro Design of ECT. (a) Exhaustive Self-Attention Block (ESAB). (b) Feed Forward Network (FFN). (c) Exhaustive Self-Attention (ESA). Key designs in ESA are the Spectral-wise Discontinuous 3D (SD3D) splitting and alignment strategies, the Dynamic Low-Rank Mapping (DLRM) model, and the Unified Spatial-Spectral self-Attention (USSA) model. }
    \vspace{-2em}
    \label{fig:esa}
 \end{figure}

\subsection{Micro Design}

Since the main difference between ESAB$\rm {_C}$ and ESAB$\rm {_I}$ lies in the subsequent processing, whether it is for correlations between tokens or within tokens, let us focus on ESAB$\rm {_C}$ to illustrate the micro design. Furthermore, from this point onward, we will no longer distinguish between ESAB$\rm {_C}$ and ESAB$\rm {_I}$ in the mathematical notation below. 

As depicted in Figure~\ref{fig:esa}, an Exhaustive Self-Attention Block (ESAB) comprises Exhaustive Self-Attention (ESA), Layer Normalization, and a Feed-Forward Network. The key process in ESA is summarized as follows: Firstly, the Spectral-wise Discontinuous 3D (SD3D) splitting strategy is applied to generate tokens, facilitating the exploitation of unified spatial-spectral correlation. Following that, the Low-Rank Dependence Map is generated through the Dynamic Low-Rank Mapping (DLRM) module to model linear dependence among multiple tokens. The Full-Rank Attention Map is generated through the Unified Spatial-Spectral self-Attention (USSA) module to model the independent correlation between pairs of tokens. Next, we introduce the implementation details of ESA.

First, the feature map undergoes two layers of grouped convolutions to learn dynamic positional encoding, which is added to the feature map to model the position of each token. Then, the feature map is linearly transformed into a hidden space. A Spectral-wise Discontinuous 3D (SD3D) splitting operation is performed to generate $Q$, $K$, and $V$. SD3D splitting strategy contains continuous splitting in the spatial dimension and discontinuous splitting in the spectral dimension, which allows for a more effective focus on spectral-wise non-local features without disrupting the continuous structure in the spatial dimension. The original feature map has dimensions $H\times W\times C$, after the SD3D splitting, the number of tokens, denoted as $n$, becomes $C\times s/c$, and the dimension of each token, denoted as $d$, becomes $H \times W \times c/s^2$, where $s$ and $c$ are hyperparameters. To simplify the expression, the multi-head attention mechanism is omitted here.

\begin{figure}[t]
   \centering
    % \includegraphics{sec/figs/fig_bubble.pdf}
 \includegraphics[width=1.0\linewidth]{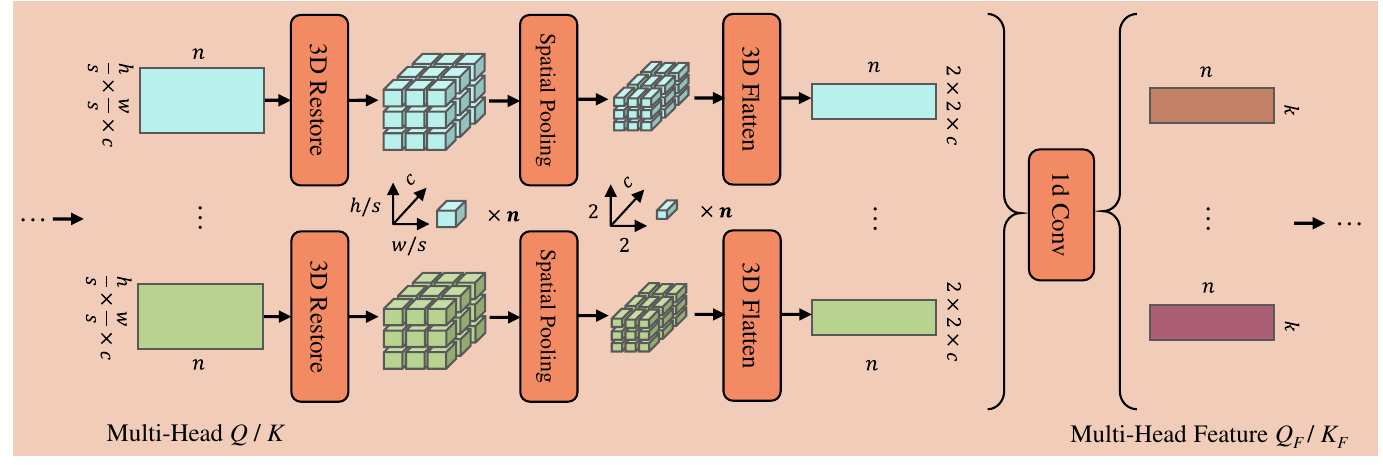}
    \caption{Detailed design of the Dynamic Low-Rank Mapping (DLRM) module.}
    \vspace{-1.6em}
    \label{fig:dlrm}
 \end{figure}

 Then, the Unified Spatial-Spectral self-Attention (USSA) is applied to capture independent full-rank correlations between pairs of tokens. The calculation of the Full-Rank Attention Map in USSA is expressed by  
\begin{equation}
    {\rm USSA}(Q, K)= \sigma \left(\tau \frac{K^T\times Q}{||K||\cdot||Q||}\right),
    \label{ussa}
\end{equation}
where $\sigma $ denotes softmax and $\tau$ is a learnable parameter. $Q=W_QX$, $K=W_KX$, and $V=W_VX$. 
${\rm L_2}$ normalization is performed within each Token to stabilize the training, and then the expressive power is improved by the learnable parameter $\tau$. It is worth noting that the $\rm L_2$ normalization and learnable parameter $\tau$ are designed by~\cite{ali2021xcit} to accommodate variable token sizes, which have been followed by abundant spectral-wise self-attention based methods~\cite{cai2022mask,cai2022mst++,zamir2022restormer,wang2023learning_lite}. We align with these designs in this paper. Since the dimension of tokens $d$ is greater than the number of tokens $n$ in this scenario and the Softmax after the dot product, there is a lower risk of rank reduction in the attention maps, as discussed in~\cite{bhojanapalli2020low}. Typically, the learned attention maps have a full rank or nearly full rank. From the optimization point of view, since the linear correlation is different in different HSIs, the loss can only be minimized when the attention mechanism learns full-rank or nearly full-rank attention maps. The experiments confirm this point as well, and we will discuss it in the supplementary material. Furthermore, the scaled dot-product attention is calculated independently between paired tokens and cannot capture the linear dependence among multiple tokens.

% \rm{Softmax}\left(conv\left(pooling\left(K\right) \right)\right)^T \rm{Softmax}\left(conv\left(pooling\left(Q\right) \right)\right) 
\begin{wrapfigure}{t}{0.51\textwidth}
	\vspace{-8mm}
	\begin{center}
		\includegraphics[width=0.51\textwidth]{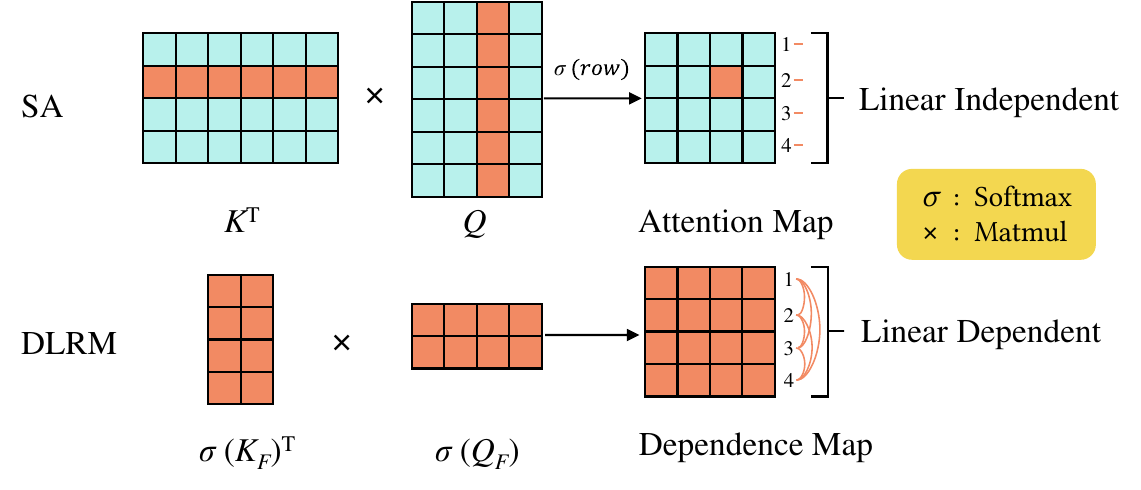}
	\end{center}
	\vspace{-3mm}
	\caption{\small Comparison of the Attention Map in Self-Attention (SA) and the Dependence Map in DLRM.}
	\vspace{-5mm}
	\label{fig:lin}
\end{wrapfigure}

To address the limitation of self-attention in modeling linear dependence within HSI, we propose a Dynamic Low-Rank Mapping (DLRM) module. 
The details of DLRM are illustrated in Figure~\ref{fig:dlrm}. Initially, the tokens from the multi-head $Q$ ($K$) are restored in a 3D manner and subsequently spatially pooled, reducing the spatial dimensions of the tokens from $h/s \times w/s$ to $2\times 2$. Following this step, the tokens are flattened in a 3D fashion to form a two-dimensional matrix. Finally, interactions take place among various heads and tokens through a 1d convolution, yielding a feature $Q_F$ ($K_F$) with dimensions $n\times k$, where $k$ is a hyperparameter and $k<n$. The matrices $Q_F$ and $K_F$ then undergo a Softmax function, followed by transposition and multiplication to generate a dynamic $n\times n$ matrix, which is a low-rank matrix with a rank no greater than $k$.  The calculation of the low-rank dependence map in DLRM is expressed by 

\vspace{-1em}
\begin{equation}
   {\rm DLRM}(Q, K)= \sigma (K_F) ^T\times \sigma (Q_F).
   \label{dlrm}
\end{equation}

The difference between the Attention Map in self-attention and the Dependence Map in DLRM is illustrated in Figure~\ref{fig:lin}. As shown in the figure, the calculation of the correlation in self-attention is token-to-token, with each row of the correlation matrix independently calculated. Each element in the matrix is solely associated with two tokens. In contrast, DLRM first facilitates information exchange among various tokens and attention heads. Therefore, each element in the Dependence Map gathers information from multiple tokens. Consequently, each element in the Dependence Map aggregates information from multiple tokens. Moreover, the Dependence Map can implicitly model the linear correlations among multiple tokens due to the low-rank nature. In summary, the Dependence Map can effectively capture the linear dependence among multiple tokens, which is absent in the Attention Map.

Then the correlations learned by USSA and DLRM are used for token fusion. First, $V$ is multiplied by the Full-Rank Attention Map learned by USSA. Following this, $V$ undergoes a linear transformation with the learnable parameter $W$ and is subsequently multiplied by the Low-Rank Dependence Map learned by DLRM.  The overall arithmetic process of ESA can be summarized by 
\begin{equation}
   {\rm ESA}(X)= {\rm DLRM}(Q,K) \times W \times {\rm USSA}(Q,K)\times V.
   \label{esa}
\end{equation}

After the token fusion, a Spectral-wise Discontinuous 3D (SD3D) alignment is performed to restore the feature map to its original shape. Finally, channel shuffling is applied to fully explore spectral-wise non-local features.

\begin{table}[t] \scriptsize
   \centering
   % \width=0.9\textwidth
   \caption{The experimental results on the NTIRE 2022~\cite{arad2022ntire} dataset, NTIRE 2020 Real World Track~\cite{arad2020ntire} dataset, and ICVL~\cite{arad2016sparse} dataset are as follows, with \textbf{bold} indicating the first place and \underline{underlined} indicating the second place.}
   \label{tab:main_exp}
   \arrayrulecolor{black}
   \resizebox{\linewidth}{!}{
   \begin{tabular}{cccccccccc} 

   % \hline
   \toprule[0.15em]
   % \Xhline{0.1em}
   \multirow{2}{*}{Method} & \multicolumn{2}{c}{NTIRE 2022} & \multicolumn{2}{c}{NTIRE 2020} & \multicolumn{2}{c}{ICVL} & \multirow{2}{*}{\begin{tabular}[c]{@{}c@{}}Params\\ (M) \end{tabular}} & \multirow{2}{*}{\begin{tabular}[c]{@{}c@{}}FLOPs\\ (G) \end{tabular}} & \multirow{2}{*}{\begin{tabular}[c]{@{}c@{}}Latency\\ (ms) \end{tabular}}  \\ 
   % \cline{2-7}
   % \cmidrule{2-7}
                           & MRAE   & RMSE                  & MRAE   & RMSE                   & MRAE   & RMSE             &                          &                         &                                                                              \\ 
   % \hline
   \midrule[0.1em]
   \vspace{0.5em}
   HSCNN+~\cite{shi2018hscnn+}                  & 0.3814 & 0.0588                 & 0.0684 & 0.0182                 & 0.2322 & 0.0424           & 4.65                    & 304.45                 &      201                                                                        \\ 
   % \arrayrulecolor{black}\cline{1-5}\cline{8-9}
   \vspace{0.5em}
   HR Net~\cite{zhao2020hierarchical}          & 0.3476 & 0.0550                 & 0.0682 & 0.0179                 &  0.1139  &  0.0313         & 31.70                   & 163.81                 &        \underline{85}                                                                      \\ 
   % \cline{1-5}\cline{8-9}
   \vspace{0.5em}
   AWAN~\cite{li2020adaptive}         & 0.2191 & 0.0349                 & 0.0668 & 0.0175                 &  0.1040 &  0.0252          & 4.04                    & 270.61                 &         312                                                                     \\

   % \cline{1-5}\cline{8-9}
   \vspace{0.5em}
   Restormer~\cite{zamir2022restormer}      & 0.1833 & 0.0274                 & 0.0645 & 0.0157                 &  0.0945  &   0.0230         & 15.11                   & 93.77                  &         178                                                                     \\ 
   % \cline{1-5}\cline{8-9}
   \vspace{0.5em}
   DRCR~\cite{li2022drcr}        & 0.1823 & 0.0288                 & 0.0664 & 0.0171                 &  0.0763   &  0.0164           & 9.48                    & 586.61                 &       300                                                                       \\

   \vspace{0.5em}
   MST++~\cite{cai2022mst++}      & 0.1645 & 0.0248                 & 0.0624 & \underline{0.0155}                 &  0.0691  &   \underline{0.0144}        & 1.62                    & 22.29                  &      112                                                                       \\ 
   % \cline{1-5}\cline{8-8}
   \vspace{0.5em}
   HySAT~\cite{wang2023learning_lite}             & \underline{0.1599} & \underline{0.0246}                 & \underline{0.0589} & \textbf{0.0142}                 &  \underline{0.0654} &    0.0154       & \underline{1.40}                    & \underline{21.08}                  &      126                                                                        \\

   \vspace{0.5em}
   \textbf{ECT (Ours)}              & \textbf{0.1564} & \textbf{0.0236}                 & \textbf{0.0588} & \textbf{0.0142}                 &  \textbf{0.0635} &    \textbf{0.0142}       & \textbf{1.19}                    & \textbf{16.75}                  &     \textbf{82}                                                                         \\
                                                              
   \bottomrule[0.15em]
   % \arrayrulecolor{black}\hline
   \end{tabular}
   }
   \end{table}

\section{Experiments}
\label{sec:exp}

\subsection{Dataset}

For spectral super-resolution experiments on simulated data, we utilized open-source datasets, including NTIRE 2022~\cite{arad2022ntire}, NTIRE 2020 Real World Track~\cite{arad2020ntire}, and ICVL~\cite{arad2016sparse}. To further validate the generalization ability of the algorithm, we conduct spectral super-resolution experiments on real RGB data.

The NTIRE 2022 dataset is currently the most complex dataset for spectral super-resolution. It is captured using the Specim IQ camera and includes a wide variety of scenes and colors. The dataset contains a total of 950 available images, including 900 training images and 50 validation images. The spatial resolution of the images is $482\times 512$, and they consist of 31 spectral channels sampled at 10$nm$ intervals, covering the wavelength range from 400$nm$ to 700$nm$.

The NTIRE 2020 dataset comprises 460 available images, including 450 training images and 10 validation images. The spatial and spectral resolutions of these images are consistent with the NTIRE 2022 dataset.

The ICVL dataset comprises 203 available HSI images. The spatial resolution is $1392\times 1300$, and the spectral resolution is consistent with the NTIRE datasets. We randomly select 20 images for the validation and others for the training.

 For the real data experiments, there is no available real dataset and most existing spectral super-resolution methods focus on fitting simulated data. Hence, we capture several real RGB images using FLIR Blackfly S BFS-U3-31S4 and obtained the spectral curves of the regions for validation using Specim IQ. We use HSI from the NTIRE 2022 training set and simulate RGB images to create paired training data.
 RGB simulation and more details about the datasets can be found in the supplementary material.

\begin{figure*}[t]
   \centering
    % \includegraphics{sec/figs/fig_bubble.pdf}
 \includegraphics[width=1.0\linewidth]{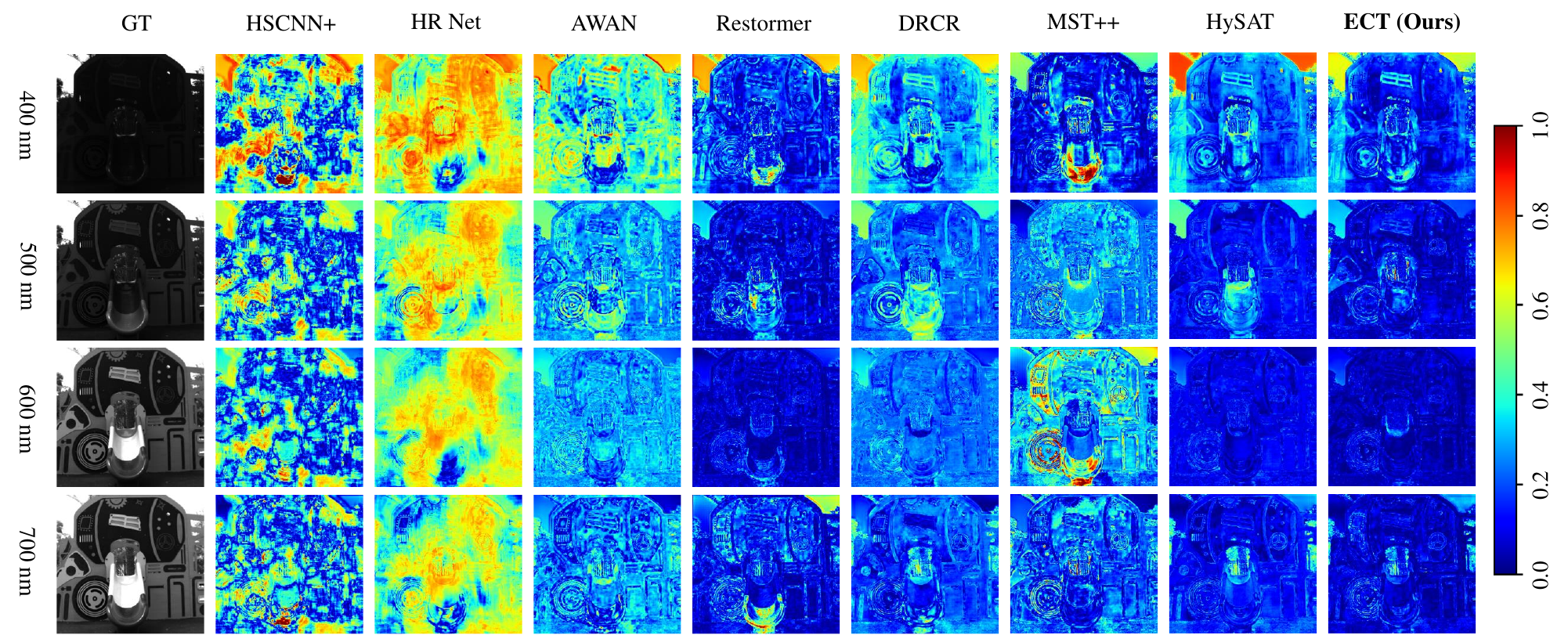}
    \caption{The MRAE heatmaps, including 400 nm, 500 nm, 600 nm, and 700 nm bands on \textit{ARAD\_0944} from the NTIRE 2022 validation data.}
    \label{fig:vis1}
 \end{figure*}

\subsection{Implementation Details}
For the hyperparameters in the network structure, we set the SD3D Splitting scale $c=4, s=2$ and low-rank factor $k=12$ for ESAB$\rm {_C}$. For ESAB$\rm {_I}$, we set the SD3D Splitting scale $c=16, s=4$ and low-rank factor $k=8$. The number of network stages $N_s$ is set to 2.

For the evaluation metrics, following the NTIRE challenges, we use the Mean Relative Absolute Error (MRAE) and Root Mean Squared Error (RMSE) metrics to evaluate the performance of each network. We primarily use MRAE as the main metric and also adopt it as the training objective. RMSE is used as an auxiliary metric.
%, defined as follows:

For the network training details, we utilize a batch size of 40 and employ a learning rate schedule that follows the cosine annealing scheme, decreasing from 4e$-$4 to 1e$-$6 over 3e5 iterations. We choose the AdamW optimizer with parameters $\beta_1 =0.9$, $\beta_2 =0.999$, $\epsilon=10^{-8}$ and weight decay is set to 1e$-$4. RGB images are first split into $128\times 128$ patches and undergo random rotations and flips for data augmentation before being input into the network. The codes and pretrained models will be made publicly available.

\begin{figure*}[t]
   \centering
    % \includegraphics{sec/figs/fig_bubble.pdf}
 \includegraphics[width=1.0\linewidth]{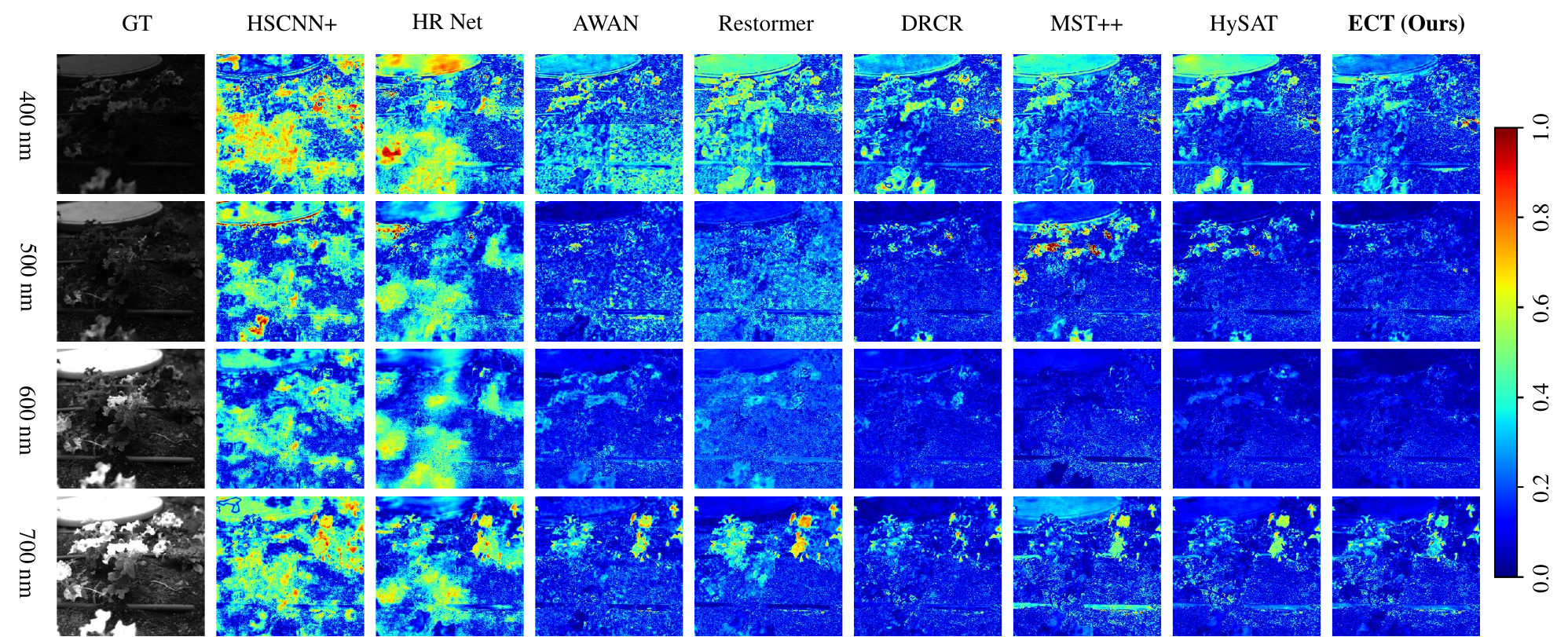}
    \caption{The MRAE heatmaps, including 400 nm, 500 nm, 600 nm, and 700 nm bands on \textit{ARAD\_0940} from the NTIRE 2022 validation data.}
    \label{fig:vis2}
 \end{figure*}

\subsection{Results on Simulated Data}
%\TODO{need to supply spectral curve comparison}
\subsubsection{Quantitative Results}
On NTIRE 2022~\cite{arad2022ntire}, NTIRE 2020~\cite{arad2020ntire}, and ICVL~\cite{arad2016sparse} datasets, we compared ECT with various neural networks, as presented in Table~\ref{tab:main_exp}. HSCNN+ is the champion in the NTIRE 2018 Clean Track and Real World Track. HR Net and AWAN are the champions in the NTIRE 2020 Clean track and Real World track, respectively.  MST++ and DRCR are the first and third place in NTIRE 2022, respectively. Restormer is an advanced algorithm in image reconstruction that has a core design similar to MST++. HySAT is the SOTA spectral super-resolution method published recently.
Among all the methods, ECT achieves the lowest MRAE with the lowest computational costs and the smallest number of parameters. We further test the inference latency of all models with the same input size $512\times512\times3$ on the same 3090 GPU. We select the average inference latency over 30 runs when the inference latency is stabilized. It is worth noting that compared to the existing SOTA HySAT, our ECT further enjoys a 34\% reduction in inference time. The results demonstrate the superior performance of our method and highlight the significance of modeling unified spatial-spectral correlation and linear dependence.

\begin{figure*}[t]
   \centering
    % \includegraphics{sec/figs/fig_bubble.pdf}
 \includegraphics[width=1.0\linewidth]{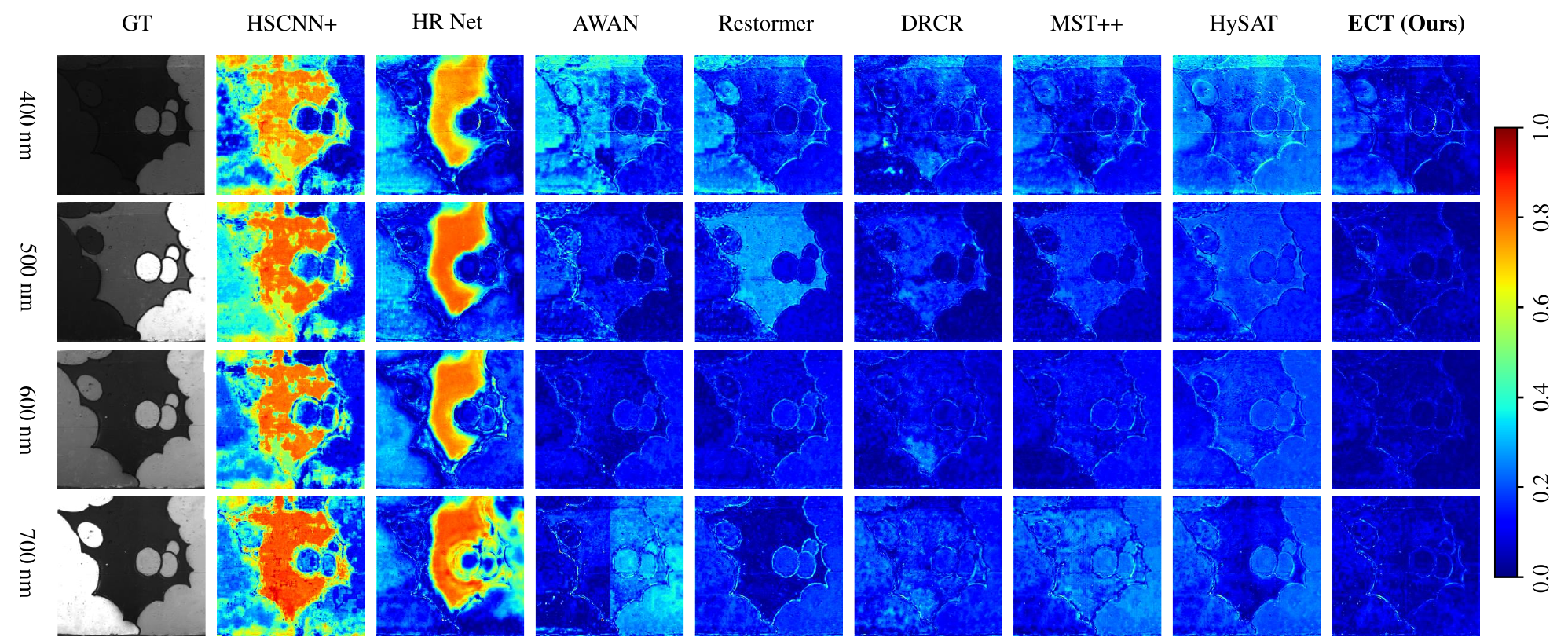}
    \caption{The MRAE heatmaps, including 400 nm, 500 nm, 600 nm, and 700 nm bands on \textit{ARAD\_0903} from the NTIRE 2022 validation data.}
    \label{fig:vis3}
 \end{figure*}

\begin{figure}[t]
   \centering
   \scalebox{0.8}{
   \subfloat{
   % \hspace{-0.198em}
   \includegraphics[width=0.3\linewidth]{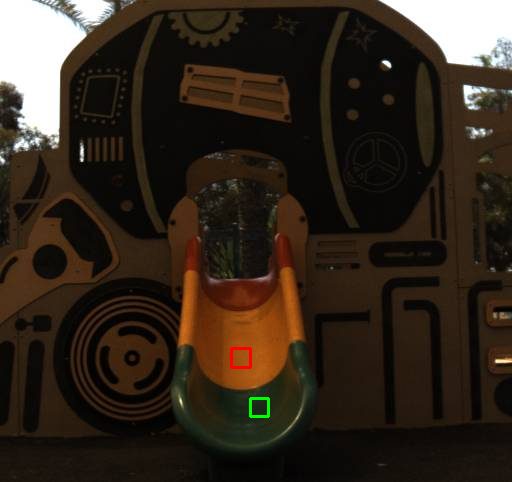}
   \vspace{0.3em}
   }      
   }
   \subfloat{
   % \hspace{-1.1em}
   \includegraphics[width=0.3\linewidth]{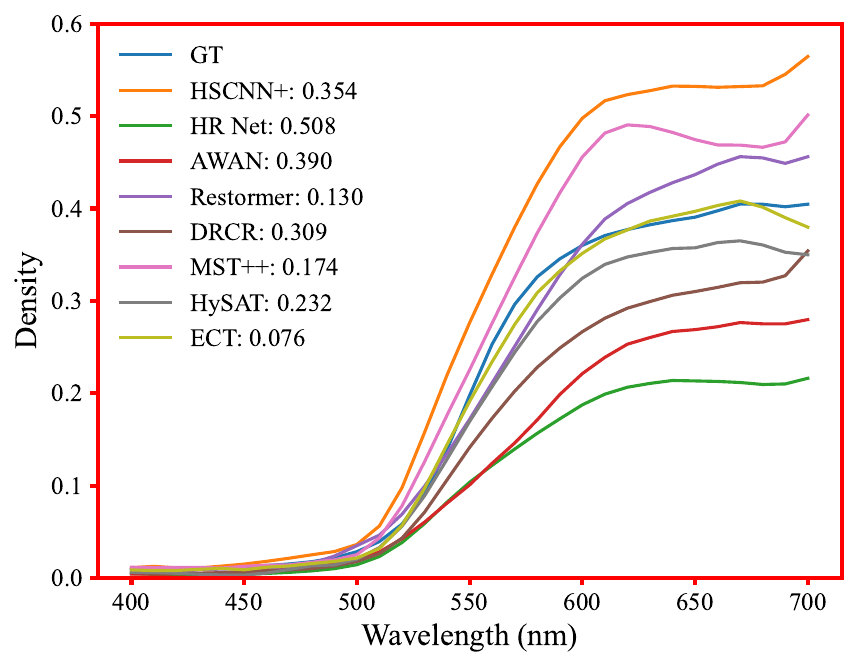}
   \vspace{-0.3em}
   }         
   \subfloat{
   % \hspace{-1.1em}
   \includegraphics[width=0.3\linewidth]{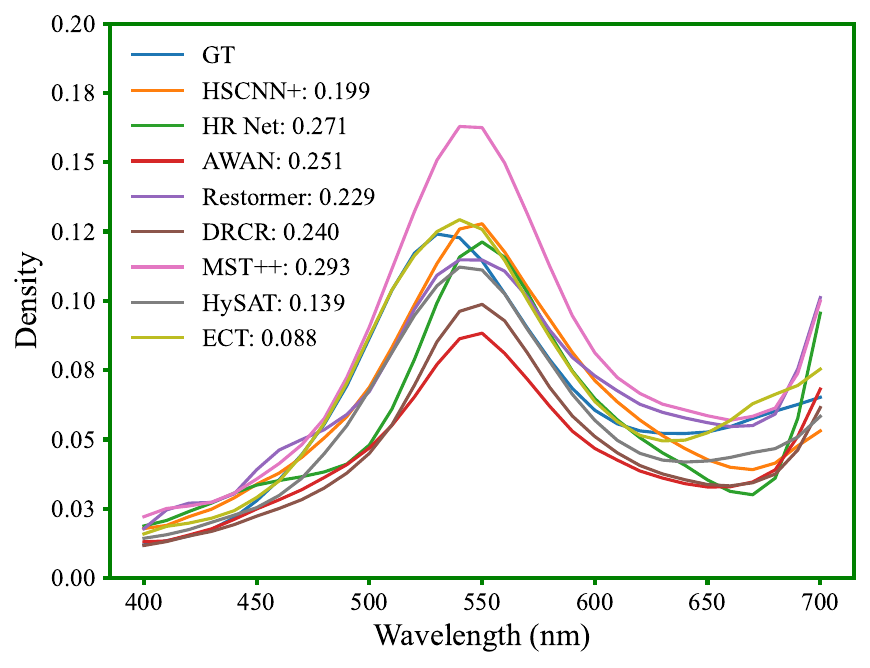}
   \vspace{-0.3em}
   }   
   \\
   \scalebox{0.8}{
   \subfloat{
   % \hspace{0.98em}
   \includegraphics[width=0.3\linewidth]{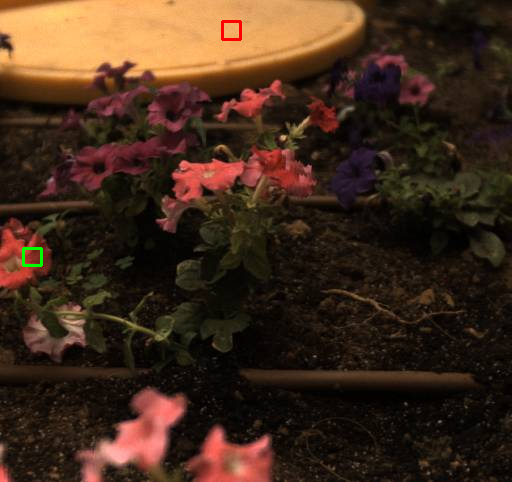}
   \vspace{0.3em}
   }      
   }
   \subfloat{
   % \hspace{-0.3em}
   \includegraphics[width=0.3\linewidth]{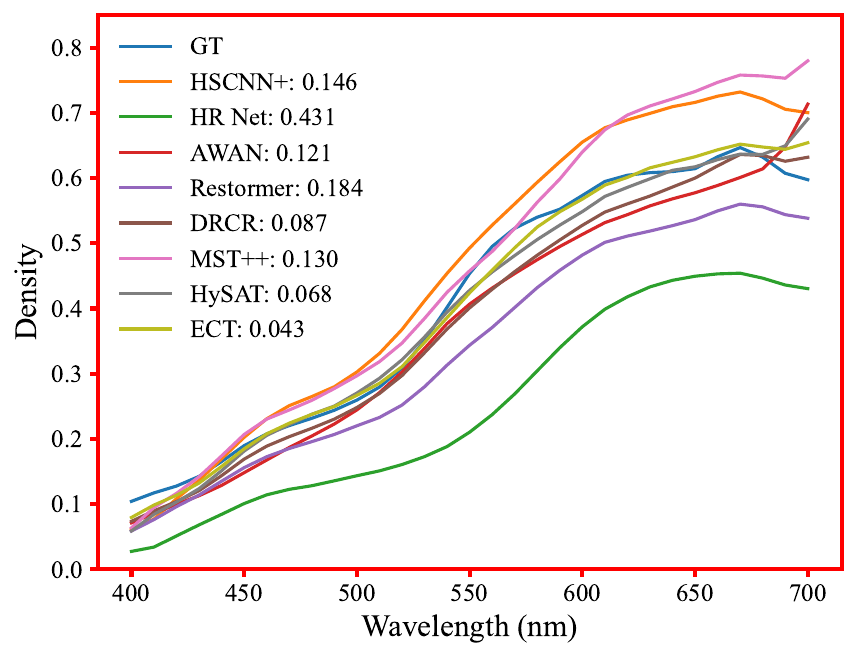}
   \vspace{-0.2em}
   }
   \subfloat{
   % \hspace{-0.3em}
   \includegraphics[width=0.3\linewidth]{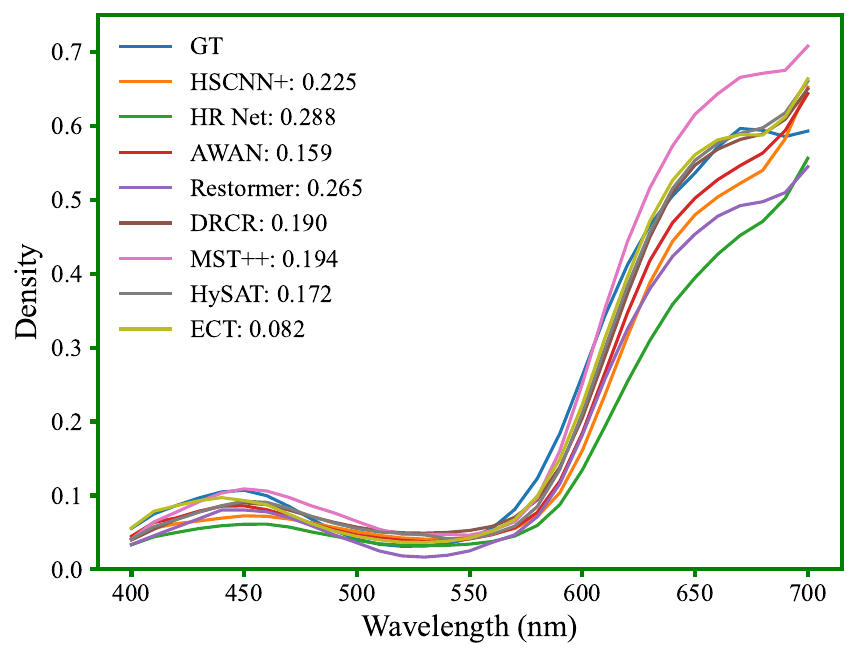}
   \vspace{-0.2em}
   }
   \\
   \scalebox{0.8}{
      \subfloat{
      % \hspace{0.98em}
      \includegraphics[width=0.3\linewidth]{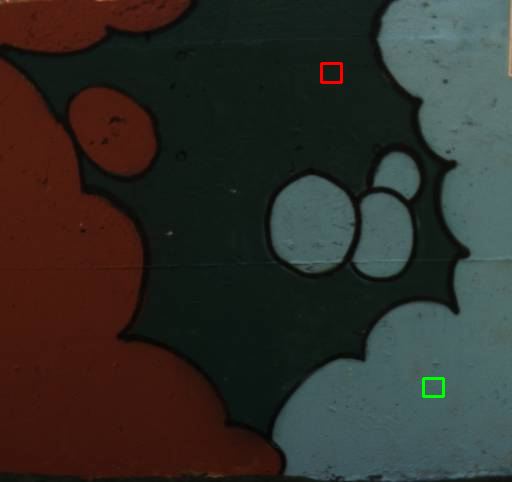}
      \vspace{-0.3em}
      }      
      }
      \subfloat{
      % \hspace{-0.3em}
      \includegraphics[width=0.3\linewidth]{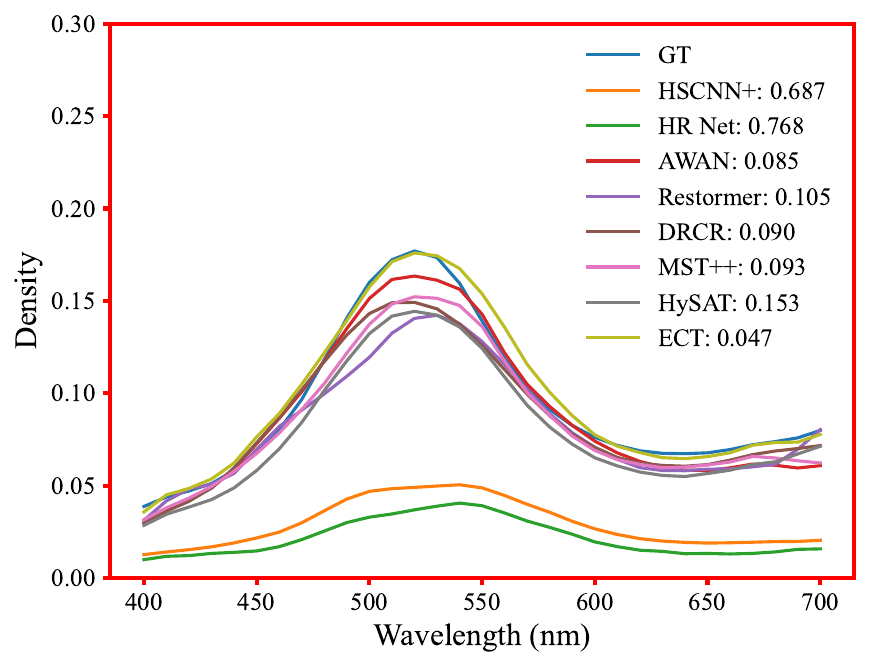}
      \vspace{-0.6em}
      }
      \subfloat{
      % \hspace{-0.3em}
      \includegraphics[width=0.3\linewidth]{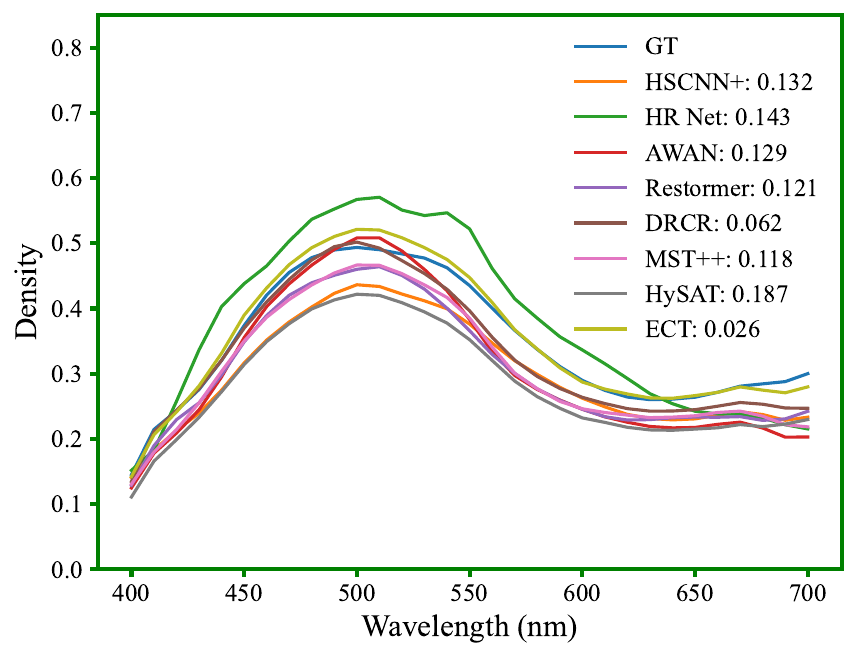}
      \vspace{-0.6em}
      }

    \caption{Comparison of reconstructed spectral curves and MRAE in the small regions. The images in the first line are the official RGB of \textit{ARAD\_0944}, \textit{ARAD\_0940} and \textit{ARAD\_0903} from the NTIRE 2022 validation data. The spectral curves with the red and green axes correspond
to the red and green boxes in the corresponding figure.}
\vspace{-3em}
    % The spectral curves with the red and green axes correspond to the red and green boxes in the figure. Please zoom in for a better view.
    \label{fig:curve1}
 \end{figure}

\hspace{-2em}
\begin{minipage}[t]{0.6\textwidth} \scriptsize
   \centering
   
   \makeatletter\def\@captype{table}
    \caption{\small Experimental results on real data.}
   \vspace{-0.8em} 
   \label{exp:real}  
   \begin{tabular}{ccccccc} 

      % \hline
      \toprule[0.1em]
      \multirow{2}{*}{Method} & \multicolumn{3}{c}{Outdoor Scene} & \multicolumn{3}{c}{Indoor Scene}  \\ 

    & MRAE & RMSE & SAM   & MRAE   & RMSE & SAM \\
    \midrule[0.05em]
    \vspace{0.2em}
           DRCR      &   0.2143   & \underline{0.0070} & 0.1896 & 0.2499  & 0.0104 &0.2340\\
           \vspace{0.2em}
        MST++        &   0.2622   & 0.0084 & 0.2245 & 0.2257  & 0.0091 & 0.2055 \\
        \vspace{0.2em}
            HySAT     &   \underline{0.2135}  & \underline{0.0070} & \underline{0.1868} & \underline{0.2202} & \underline{0.0088}&\underline{0.1974} \\
            \vspace{0em}
         \textbf{ECT}       &  \textbf{0.2012}   & \textbf{0.0065} & \textbf{0.1730} & \textbf{0.2114} & \textbf{0.0082} & \textbf{0.1831} \\ 
         % \vspace{0em}
         % \textbf{ECT}       &  \textbf{0.2120}   & \textbf{0.0065} & \textbf{0.1744} & \textbf{0.2095} & \textbf{0.0081} & \textbf{0.1818} \\ 
   \bottomrule[0.1em]
   \end{tabular}      
   \end{minipage}
   \hspace{0.1em}
   \begin{minipage}[t]{0.37\textwidth} \scriptsize
      \centering
      \makeatletter\def\@captype{table}
       \caption{\small Ablation study of $N_s$.}
      \label{ab:stage}  
      %\vspace{0.6em}
      \begin{tabular}{cccccc} 
         % \hline
         \toprule[0.1em]
         % $N_s$ & MRAE   & RMSE   & Params & FLOPs & Latency \\
         % \midrule[0.05em]
         % \vspace{0.3em}
         % 1    & 0.1648 & 0.0243 & 0.60 M & 8.69 G & 41 ms \\
         % \vspace{0.3em}
         % 2       &   0.1564   & 0.0236 & 1.19 M & 16.75 G   \\
         %  \vspace{0.3em}
         %    3     &  \textbf{0.1542} & \textbf{0.0231} & 1.78 M & 24.81 G \\ 
         %    \vspace{0em}
         %    4     &  \underline{0.1554} & \underline{0.0234} & 2.37 M & 32.88 G \\             
         $N_s$ & MRAE   & RMSE   & Params  & Latency \\
         \midrule[0.05em]
         \vspace{0.3em}
         1    & 0.1648 & 0.0243 & 0.60 M  & 41 ms \\
         \vspace{0.3em}
         2       &   0.1564   & 0.0236 & 1.19 M & 82 ms   \\
          \vspace{0em}
            3     &  0.1542 & 0.0231 & 1.78 M & 124 ms \\ 
            % \vspace{0em}
            % 4     &  \underline{0.1554} & \underline{0.0234} & 2.37 M & 32.88 G \\             
      \bottomrule[0.1em]
      \end{tabular}      
      \end{minipage}
\vspace{1em}

\subsubsection{Qualitative Results}

We showcase the visual effects of the MRAE heatmaps in Figure~\ref{fig:vis1}, Figure~\ref{fig:vis2}, and Figure~\ref{fig:vis3}. We also present a comparison of spectral curves in small regions among the Ground Truth and various reconstruction algorithms in Figure~\ref{fig:curve1}. The visual results indicate that ECT exhibits the best reconstruction performance across different wavelengths, which further confirms the effectiveness of our approach.
%More qualitative results are available in the supplementary material.

\subsection{Results on Real Data}

\begin{figure}[t]
   \centering
   \subfloat{
   % \hspace{0.98em}
   \includegraphics[width=0.3\linewidth]{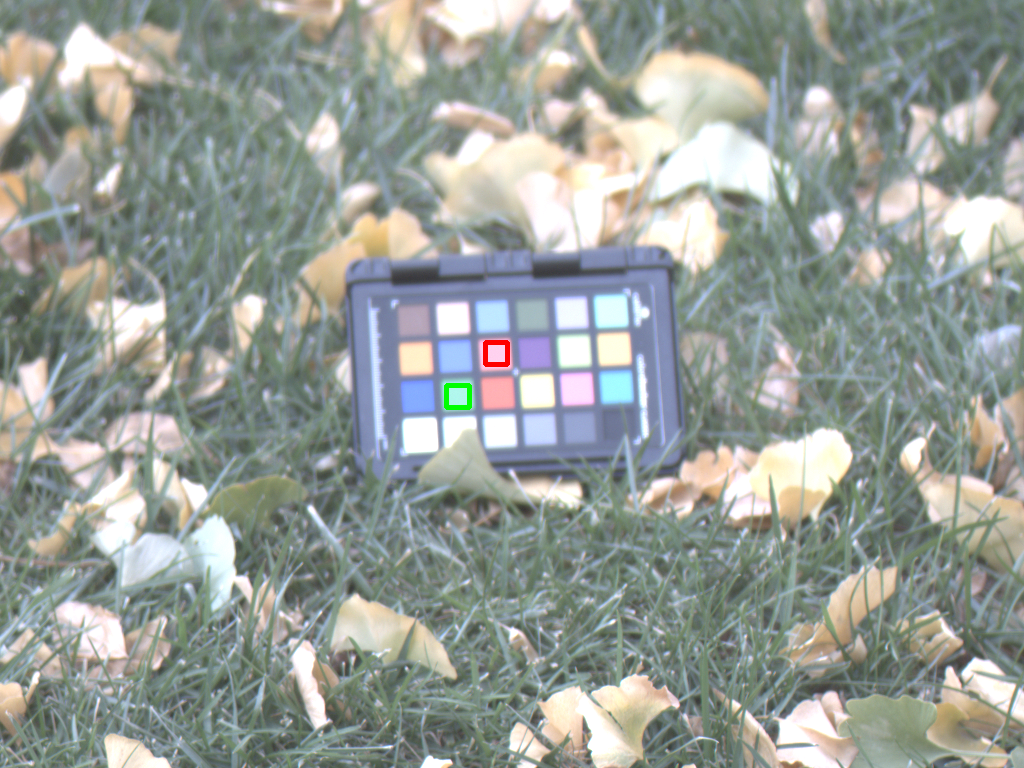}
   }
   \subfloat{
   % \hspace{-0.3em}
   \includegraphics[width=0.3\linewidth]{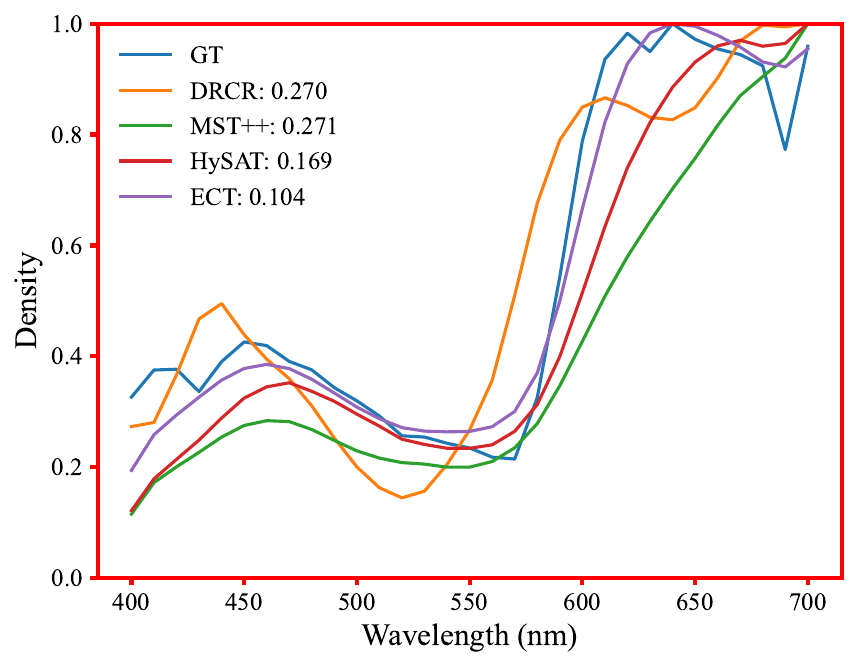}
   \vspace{-0.3em}
   }
   \subfloat{
   % \hspace{-0.3em}
   \includegraphics[width=0.3\linewidth]{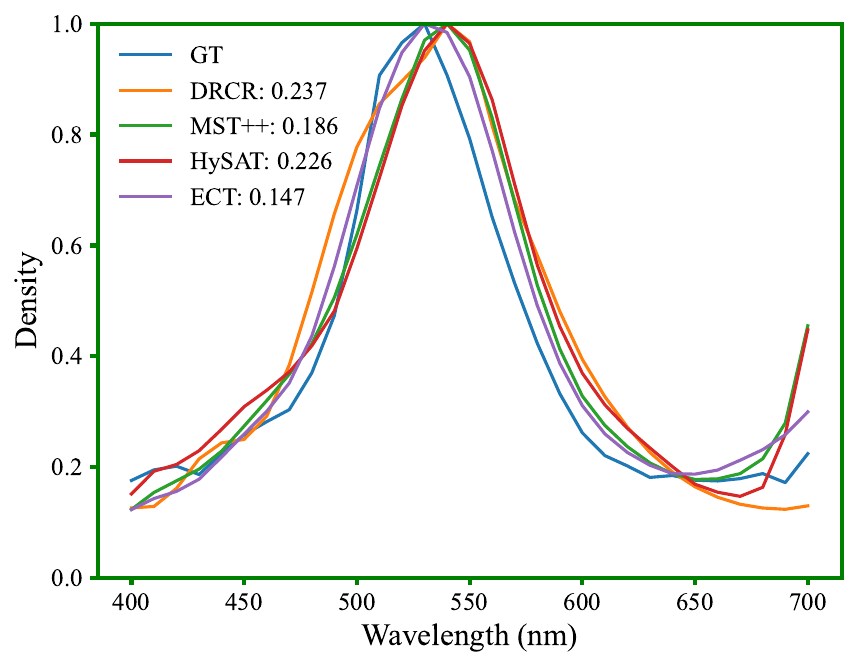}
   \vspace{-0.3em}
   }
   \\
   \subfloat{
   %  \hspace{-0.198em}
   \includegraphics[width=0.3\linewidth]{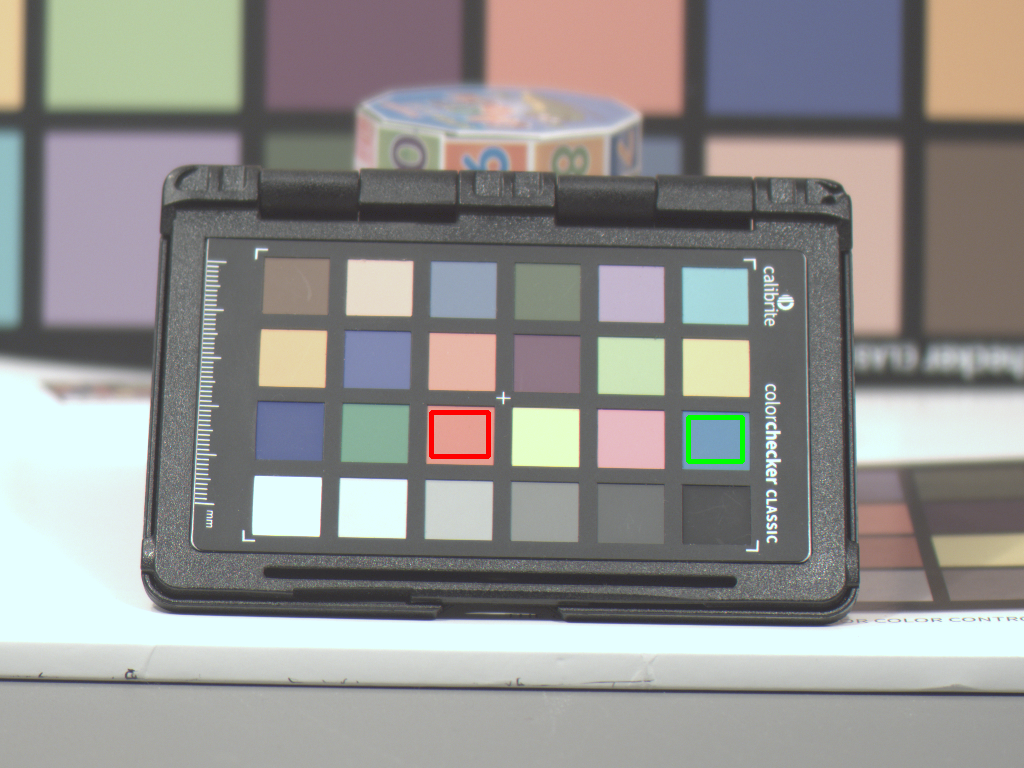}
   }
   \subfloat{
   % \hspace{-1.1em}
   \includegraphics[width=0.3\linewidth]{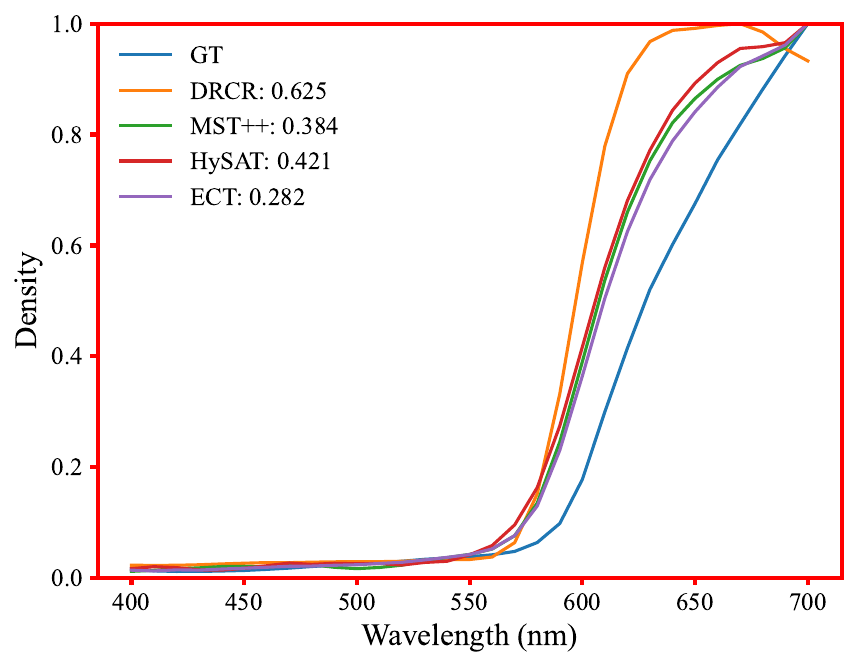}
   \vspace{-0.3em}
   }  
   \subfloat{
   % \hspace{-1.1em}
   \includegraphics[width=0.3\linewidth]{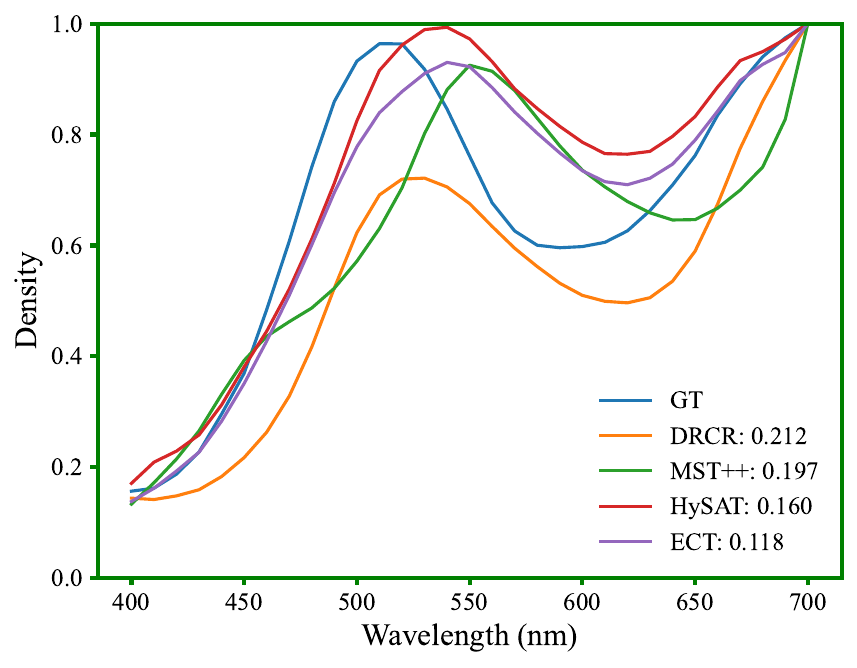}
   \vspace{-0.3em}
   }

    \caption{Comparison of reconstructed spectral curves and MRAE in the small regions. Each region is normalized to remove the influence of brightness. The images in the first line are the real RGBs captured outdoors and indoors.}
    \label{fig:curve2}
 \end{figure}

To further investigate the generalization ability of ECT, we conduct some experiments on real RGB data. We choose flattened regions for validation to avoid the influence of misalignment. We capture RGB images of a color chart both indoors under halogen lights and outdoors under sunlight, along with corresponding HSIs. We evaluated the algorithms using the average error of the 18 color patches on the color chart. We further evaluate Spectral Angle Mapper (SAM) on real data. The quantitative and qualitative results of ECT compared with three advanced spectral super-resolution methods MST++~\cite{cai2022mst++}, DRCR~\cite{li2022drcr} and HySAT~\cite{wang2023learning_lite} are shown in Table~\ref{exp:real} and Figure~\ref{fig:curve2}. The experimental results on real data demonstrate the advanced performance and the strong generalization ability of ECT. 

 \hspace{-1em}
\begin{minipage}[t]{0.48\textwidth} \scriptsize
   \centering
   
   \makeatletter\def\@captype{table}
    \caption{\small Ablation study of SD3D splitting strategy and the DLRM module.}
   \vspace{-0.8em} 
   \label{ab:main}  
   \begin{tabular}{cccccc} 

      % \hline
      \toprule[0.1em]
    SD3D & DLRM & MRAE   & RMSE   & Params & FLOPs \\
    \midrule[0.05em]
    \vspace{0.2em}
           \XSolidBrush      &   \XSolidBrush   & 0.1761 & 0.0266 & 0.55 M  & 7.84 G \\
           \vspace{0.2em}
        \Checkmark        &   \XSolidBrush   & \underline{0.1700} & \underline{0.0255} & 0.58 M  & 8.24 G \\
        \vspace{0.2em}
            \XSolidBrush     & \Checkmark    & 0.1733 & 0.0261 & 0.56 M  & 8.27 G \\
            \vspace{0em}
         \Checkmark       &  \Checkmark   & \textbf{0.1648} & \textbf{0.0243} & 0.60 M  & 8.69 G \\ 
   \bottomrule[0.1em]
   \end{tabular}      
   \end{minipage}
   \hspace{0.2em}
   \begin{minipage}[t]{0.48\textwidth} \scriptsize
      \centering
      \makeatletter\def\@captype{table}
       \caption{\small Ablation study of the token splitting strategy.}
      \label{ab:split}  
      %\vspace{0.6em}
      \begin{tabular}{ccccc} 
         % \hline
         \toprule[0.1em]
         Splitting & MRAE   & RMSE   & Params & FLOPs \\
         \midrule[0.05em]
         \vspace{0.3em}
         Spectral-wise    & \underline{0.1740} & \underline{0.0257} & 0.59 M & 8.69 G \\
         \vspace{0.3em}
         Spatial-wise       &   0.1937   & 0.0285 & 0.94 M & 14.35 G   \\

          \vspace{0em}
            SD3D      &  \textbf{0.1648} & \textbf{0.0243} & 0.60 M & 8.69 G \\          
      \bottomrule[0.1em]
      \vspace{0.4em}
      \end{tabular}      
      \end{minipage}

\hspace{-1em}  
   \begin{minipage}[t]{0.48\textwidth} \scriptsize
      \centering
      \makeatletter\def\@captype{table}
       \caption{\small Ablation study of the continuity in SD3D splitting strategy.}
      \label{ab:scale}  

      \begin{tabular}{cccc} 
         % \hline
         \toprule[0.1em]
         Spatial-wise & Spectral-wise & MRAE$\downarrow$   & RMSE$\downarrow$    \\
         \midrule[0.05em]
         \vspace{0.3em}
         Continuous         &  Continuous    & 0.1769 & \underline{0.0263}   \\
         \vspace{0.3em}
         Continuous       &   Discontinuous   & \textbf{0.1648} & \textbf{0.0243}   \\

          \vspace{0em}
            Discontinuous      &  Discontinuous & \underline{0.1739} & \underline{0.0263} \\
      \bottomrule[0.1em]
      \vspace{-2em}
      \end{tabular}      
      \end{minipage}
      \begin{minipage}[t]{0.48\textwidth} \scriptsize
         \centering
         \makeatletter\def\@captype{table}
         \caption{Ablation study of the low-rank factor $k$.}
         \vspace{-0.5em}
         \label{ab:fac}  
         \begin{tabular}{ccccc} 
            % \hline
            \toprule[0.1em]
            Factor $k$ & MRAE   & RMSE   & Params & FLOPs \\
            \midrule[0.05em]
            \vspace{0.2em}
            8    & 0.1689 & 0.0260 & 0.60 M & 8.69 G \\
            \vspace{0.2em}
            12       &   \textbf{0.1648}   & \textbf{0.0243} & 0.60 M & 8.69 G   \\

             \vspace{0.2em}
               16      &  \underline{0.1671} & \underline{0.0249} & 0.60 M & 8.69 G \\
               \vspace{0em}  
               32      &  0.1701 & 0.0259 & 0.62 M & 8.69 G \\           
         \bottomrule[0.1em]
         \vspace{-2em}
         \end{tabular}      
         \end{minipage}

\subsection{Ablation Study}

To fully explore the effects and the working mechanics of the whole architecture, Spectral-wise Discontinuous 3D (SD3D) splitting strategy, and the Dynamic Low-Rank Mapping (DLRM) module, we introduce some ablation studies here. More ablation studies can be found in the supplementary material. All ablation studies are conducted on the NTIRE 2022 dataset.

\subsubsection{Ablation Study of the Network Stage $N_s$}

We conduct an ablation study on the number of stages ($N_s$) in the network, and the results are shown in Table~\ref{ab:stage}. We primarily set $N_s=2$ balancing performance and inference latency. All ablation studies below are conducted using a 1-stage structure to efficiently validate the effectiveness of each module. 

\subsubsection{Ablation Study of SD3D Splitting and DLRM}

The Spectral-wise Discontinuous 3D (SD3D) splitting strategy and the Dynamic Low-Rank Mapping (DLRM) are our two significant contributions. The SD3D splitting strategy is used to model unified spatial-spectral correlation, while DLRM is employed to capture linear dependence. We test the performance improvement of SD3D and DLRM compared to MST++~\cite{cai2022mst++} without these two structures. The experimental results, as shown in Table~\ref{ab:main}, demonstrate that both the SD3D splitting strategy and DLRM can lead to improvements. When used together, they achieve even greater performance improvements. 
The results indicate the effectiveness of our two key designs.

\subsubsection{Ablation Study of SD3D Splitting Strategy}

First, We compare our SD3D splitting strategy with traditional spatial-wise and spectral-wise splitting strategies. The results shown in Table~\ref{ab:split} demonstrate the effectiveness of our method. Moreover, the key characteristic of the SD3D splitting strategy lies in its spatial-wise continuity and spectral-wise discontinuity splitting approach. Discontinuous splitting allows for a greater focus on non-local information, while continuous splitting helps preserve the local structure. We conducted an ablation study on the continuity and discontinuity in both spectral and spatial directions, as shown in Table~\ref{ab:scale}. Experiments indicate that spectral super-resolution benefits from non-local features in the spectral direction, and adverse effects arise when disrupting spatial continuity.

\subsubsection{Ablation Study of Low-Rank Factor $k$}

The critical parameter in the DLRM module is the number of feature $Q_F$ ($K_F$) columns, denoted as $k$. The rank of the low-rank dependence map in DLRM is not greater than $k$. The experimental results for different values of $k$ in ESAB$\rm {_C}$ are shown in Table~\ref{ab:fac}. When $k = 32$, it means $k = n$, which does not constrain the dependence map to be low-rank. The experimental results highlight the importance of the low-rank characteristic of the dependence map.

\begin{table}[t] \scriptsize
   \centering
   % \width=0.9\textwidth
   \caption{Comparison with the SOTA method of CASSI-based spectral reconstruction.}
   \label{tab:cassi_cmp}
   \arrayrulecolor{black}
   \resizebox{\linewidth}{!}{
   \begin{tabular}{cccccccccccc} 

   % \hline
   \toprule[0.1em]

   \multirow{2}{*}{Method} & \multicolumn{3}{c}{NTIRE 2022} & \multicolumn{3}{c}{NTIRE 2020} & \multicolumn{3}{c}{ICVL} & \multirow{2}{*}{\begin{tabular}[c]{@{}c@{}}Params\\ (M) \end{tabular}} & \multirow{2}{*}{\begin{tabular}[c]{@{}c@{}}FLOPs\\ (G) \end{tabular}}   \\ 

                           & MRAE   & RMSE     & PSNR              & MRAE   & RMSE  & PSNR                  & MRAE   &RMSE &  PSNR      &  &                                                                                \\ 
   % \hline
   \midrule[0.06em]
  \vspace{-0.05em}
   % \midrule[0.1em]
   PADUT      & 0.1850 & 0.0271     &      33.04      & 0.0624 & 0.0158      &   37.07        & 0.0798   &   0.0193             &    36.77    & 1.71 &              20.29                         \\                                                       

   \textbf{ECT}              & \textbf{0.1564} & \textbf{0.0236}     &\textbf{34.81}            & \textbf{0.0588} & \textbf{0.0142}    &     \textbf{37.71}       &  \textbf{0.0635} &    \textbf{0.0142}  &\textbf{38.50}          & \textbf{1.19} &   \textbf{16.75}                                                      \\
   
   \bottomrule[0.1em]
   \vspace{-3em}
   % \arrayrulecolor{black}\hline
   \end{tabular}
   }
   \end{table}
\section{Discussion}
\label{sec:discussion}

Recently, spectral reconstruction algorithms based on CASSI have received wide interest. We find that performance improvements brought by recent advanced CASSI-based spectral reconstruction algorithms~\cite{cai2022coarse,dong2023residual,li2023pixel} are typically reflected in sharper spatial reconstructions. However, spatial details reconstruction is not a necessary challenge in the spectral super-resolution task. Therefore, recent algorithms used to improve CASSI-based spectral reconstruction have difficulty enhancing the performance of spectral super-resolution. We retrain the SOTA algorithm for CASSI-based spectral reconstruction PADUT~\cite{li2023pixel} for the spectral super-resolution task. The experimental results shown in Table~\ref{tab:cassi_cmp} demonstrate that our ECT achieves significantly better performance.

\section{Conclusion}
\label{sec:conclusion}
In this paper, we analyze the limitations of existing spectral super-resolution Transformers in modeling unified spatial-spectral correlation and linear dependence. We propose a novel Exhaustive Correlation Transformer (ECT) to model these correlations for spectral super-resolution. Specifically, we propose a Spectral-wise Discontinuous 3D (SD3D) splitting strategy to model unified spatial-spectral correlation and a Dynamic Low-Rank Mapping (DLRM) module to capture linear dependence. Experimental results demonstrate that our approach achieves state-of-the-art performance on both simulated and real data with a 34\% reduction in inference latency. In future work, we will commit to applying our method to other tasks related to hyperspectral imaging.

\bibliographystyle{splncs04}
\bibliography{main}
\end{document}